\documentclass[twocolumn,preprintnumbers,amsmath,amssymb]{revtex4}

\usepackage{graphicx}              
\usepackage{dcolumn}               
\usepackage{longtable}             
\usepackage{bm}                    

\newcommand{\D} [2]{\frac{\partial #1}{\partial #2}}

\def\beq{\begin{equation}}
\def\eeq{\end{equation}}

\begin{document}

\title{Compact and Flexible Basis Functions for Quantum Monte Carlo Calculations}

\author{F. R. Petruzielo$^{1}$}
\email{frp3@cornell.edu}
\author{Julien Toulouse$^{2,3}$}
\email{julien.toulouse@upmc.fr}
\author{C. J. Umrigar$^{1}$}
\email{CyrusUmrigar@cornell.edu}
\affiliation{
$^1$Laboratory of Atomic and Solid State Physics, Cornell University, Ithaca, New York 14853, USA.\\
$^2$Laboratoire de Chimie Th\'eorique - UMR 7616, Universit\'e Pierre et Marie Curie (UPMC Univ Paris 06), 75005 Paris, France.\\
$^3$Laboratoire de Chimie Th\'eorique - UMR 7616, Centre National de la Recherche Scientifique (CNRS), 75005 Paris, France.\\
}

\date{\today}
\begin{abstract}
Molecular calculations in quantum Monte Carlo frequently employ a mixed basis consisting of contracted and primitive Gaussian functions.
While standard basis sets of varying size and accuracy are available in the literature, we demonstrate that reoptimizing the primitive function exponents within quantum Monte Carlo yields more compact basis sets for a given accuracy.
Particularly large gains are achieved for highly excited states.
For calculations requiring non-diverging pseudopotentials, we introduce Gauss-Slater basis functions that behave as Gaussians at short distances
and Slaters at long distances.
These basis functions further improve the energy and fluctuations of the local energy for a given basis size.
Gains achieved by exponent optimization and Gauss-Slater basis use are exemplified by calculations
for the ground state of carbon, the lowest lying excited states of carbon with $^5S^o$, $^3P^o$, $^1D^o$, $^3F^o$ symmetries, carbon dimer, and naphthalene.
Basis size reduction enables quantum Monte Carlo treatment of larger molecules at high accuracy.
\end{abstract}
\maketitle
\section{Introduction}
\label{sec:intro}
In traditional quantum chemistry (QC) calculations, molecular orbitals are often expanded in a combination of contracted Gaussian basis functions and primitive Gaussian basis functions.
For each occupied orbital, a contracted function is constructed to reproduce the corresponding atomic orbital from an effectively single-electron theory such as Hartree-Fock (HF) \cite{Hariharan1973,Hehre1969},
or the natural orbital from a post-HF method \cite{DunningJr1989,Kendall1992}.

While a single primitive Gaussian has incorrect long-range asymptotic behavior, a contracted basis function can reproduce the correct asymptotics over a reasonable range.
However, even contracted functions are unable to produce the correct electron-nucleus cusps \cite{Kato1957} since they have zero gradient at the origin.
Despite these shortcomings, Gaussians are used in QC calculations because they permit analytical evaluation of the two-electron integrals \cite{Boys1950}.

In contrast to traditional QC methods, quantum Monte Carlo (QMC) calculations \cite{Foulkes2001} enjoy greater wavefunction flexibility by using Monte Carlo integration to evaluate matrix elements.
In particular, bases need not be restricted to Gaussians.
For calculations employing a potential that diverges at the nucleus, Slater basis functions can exactly reproduce the correct electron-nucleus cusp and long-range asymptotic behavior of the orbitals.
In fact, for all-electron QMC calculations, highly accurate results have been obtained by employing compact basis sets consisting of Slater functions with optimized exponents \cite{Umrigar1988,Filippi1996}.

Conversely, the basis sets used for non-divergent pseudopotential calculations in QMC have deviated little from typical QC basis sets.
For these pseudopotentials, orbitals have no electron-nucleus cusp.
In this case, Gaussian basis functions are more appropriate than Slater functions at small electron-nuclear distances but still have incorrect long-range asymptotics.

Contracted and primitive Gaussian functions are frequently splined on a radial grid for QMC.
Splining contracted Gaussians presents a definite computational advantage since evaluating polynomials is much cheaper than evaluating a linear combination of Gaussians.
In contrast, splining primitive Gaussians provides minimal benefit at best.

We propose two ideas for improving basis sets for pseudopotential calculations in QMC.
First, primitive basis function exponents are optimized for each system.
This provides greater accuracy with a compact basis for a wide range of chemical environments and excitation levels.
To facilitate optimization, the primitive basis functions remain analytic while the contracted functions are splined.

Second, we propose a novel form of primitive basis function appropriate for calculations involving non-diverging pseudopotentials.
These primitives, which we call Gauss-Slater (GS) functions, have the short-range behavior of a Gaussian function and the long-range behavior of a Slater function.

The utility of our improvements is demonstrated by calculations for carbon, the lowest lying excited states of carbon with $^5S^o$, $^3P^o$, $^1D^o$, $^3F^o$ symmetries, carbon dimer, and naphthalene.

This paper is organized as follows.
In Section \ref{sec:gs}, the form and properties of Gauss-Slater functions are introduced.
In Section \ref{sec:app}, results of our calculations are discussed.
In Section \ref{sec:conc}, concluding remarks are provided.
In the Appendices, technical details are discussed.
\section{Gauss-Slater Basis Functions}
\label{sec:gs}
We define Gauss-Slater (GS) functions as
\beq
 \varphi^{\zeta}_{nlm}(r,\theta,\phi) =N_n^\zeta \; r^{n-1} e^{-\frac{(\zeta r)^2}{1+\zeta r}} \; Z_l^m(\theta,\phi),
\eeq
where $r,\theta,\phi$ are the standard spherical coordinates, $n$ is the principal quantum number, $l$ is the azimuthal quantum number,
$m$ is the magnetic quantum number, $N_n^\zeta$ is the normalization factor, and $Z_l^m(\theta,\phi)$ is a real spherical harmonic.

Notice that for $r\ll 1$ the GS behaves like a Gaussian:
\beq
\varphi^{\zeta}_{nlm}(r,\theta,\phi) \cong N_n^\zeta \; r^{n-1} e^{-(\zeta r)^2} \; Z_l^m(\theta,\phi),
\eeq
and for $r\gg 1$ the GS behaves like a Slater:
\beq
\varphi^{\zeta}_{nlm}(r,\theta,\phi) \cong N_n^\zeta \; r^{n-1} e^{-\zeta r} \; Z_l^m(\theta,\phi).
\eeq
The GS drift velocity and local energy are well behaved at long distances, while for Gaussians these quantities diverge.

Unlike Gaussians and Slaters, normalization of GSs has no closed form expression.
Nevertheless, normalizing an arbitrary GS is trivial with the following scaling relation (see Appendix \ref{sec:scaling}) between $N_n^\zeta$ and $N_n^1$,
\beq
N_n^\zeta = \zeta^{n+1/2} \; N_n^1.
\eeq

Since GSs are not analytically integrable, the exponential part must be expanded in Gaussians for use
in quantum chemistry programs that employ analytic integrals for evaluating the matrix elements.
This expansion is
\beq
N_n^\zeta \; e^{-\frac{(\zeta r)^2}{1+\zeta r}} = \sum_i c_i^\zeta \;
\sqrt{\frac{2(2\alpha_i^\zeta)^{n+\frac{1}{2}}}{\Gamma(n+\frac{1}{2})}} \; e^{-\alpha_i^\zeta r^2},
\eeq
where $c_i^\zeta$ is the $i^{\rm{th}}$ expansion coefficient, and $\alpha_i^\zeta$ is the $i^{\rm{th}}$ Gaussian exponent.
Additionally, the following scaling relations (see Appendix \ref{sec:scaling}) hold for the expansion coefficients and Gaussian exponents:
\begin{align}
\alpha_i^\zeta &= \zeta^2 \alpha_i^1 \\
c_i^\zeta &= c_i^1.
\end{align}
Once the Gaussian expansions are found for unit exponents, expansions of arbitrary GSs follow immediately from the scaling relations.
\section{Results}
\label{sec:app}
For all applications discussed in this paper, variational Monte Carlo (VMC) and diffusion Monte Carlo (DMC) \cite{Umrigar1993} calculations are performed with the CHAMP QMC code \cite{Cha-PROG-XX} and employ the pseudopotentials and accompanying basis sets of Burkatzki, Filippi and Dolg (BFD) \cite{Burkatzki2007}.
We choose these pseudopotentials and basis functions since they were constructed for use in QMC and have proved to be quite accurate.

The wavefunction is of the standard Slater-Jastrow form.
All wavefunction parameters including Jastrow parameters, Configuration State Function (CSF) coefficients (where applicable), orbital coefficients, and primitive exponents (where applicable) are optimized via the linear method \cite{Toulouse2007,Toulouse2008,Umrigar2007}.
Optimization is performed on a linear combination of the energy and variance of the local energy with weights $0.95$ and $0.05$, respectively.
Optimizing just the energy yields slightly lower energies and somewhat higher variances.

For each system considered, calculations are performed with three different basis sets:
(1) the BFD basis, (2) fixed contracted functions and analytical Gaussian primitives with optimized exponents, and (3) fixed contracted functions and analytical Gauss-Slater primitives with optimized exponents.
We refer to these basis sets as BFD, G, and GS, respectively.

These three cases allow us to evaluate the improvements our two methods provide to the current basis sets used in QMC.
First, if both the G and GS basis sets significantly outperform the BFD basis, then the utility of reoptimizing the basis exponents within QMC will be established.
Second, the utility of the GS basis depends on its performance relative to the G basis.

\subsection{Ground State Carbon Atom}
\label{sec:c_atom}
For the carbon atom ground state, $^3P$, we consider a complete active space (CAS) wavefunction with an active space generated by distributing the four valence electrons among the thirteen orbitals of the $n=2$ and $n=3$ shells.
Denoted by CAS(4,13), this wavefunction consists of 83 CSFs comprised of 422 determinants.

In general, a single Slater determinant will not be a CSF when a certain number of electrons have been excited relative to the ground state HF Slater determinant.
However, a CSF can be produced from an arbitrary Slater determinant by applying projection operators for angular momentum $\hat{L}$ and spin $\hat{S}$.
Since states with the same $L$ and $S$ but different $L_z$ and $ S_z$ are degenerate,
we are free to choose convenient $L_z$ and $ S_z$ states.
We choose $L_z=0$ to make the wavefunctions real to within a position independent phase, and we choose $S_z=S$ to yield the minimum
number of determinants in the CSF.
Since the carbon ground state has $L=1$, $S=1$, the projection operators are of the form
\begin{align}
\hat{P}_L &= \prod_{L' \ne 1} \left[\hat{L} - L'(L'+1)\right]\\
\hat{P}_S &= \prod_{S' \ne 1} \left[\hat{S} - S'(S'+1)\right],
\end{align}
where the product over all possible angular momentum and spin values omits the desired $L=1$ and $S=1$ values.

Carbon atom VMC results for each basis set are shown in Table \ref{tab:c_atom}.
Included for comparison, coupled cluster calculations with single and double excitations and perturbative triple excitations (CCSD(T)) values for the BFD basis \cite{Burkatzki2007} exhibit much larger dependence on basis size than QMC results.

Both the G and GS basis sets outperform the BFD basis set.
The $2z$ G basis exhibits a modest gain of 0.3 mH in energy compared to the corresponding BFD basis.
The $2z$ GS basis exhibits larger gains of 1 mH in energy and 28 mH in $\sigma$, the root-mean-square (RMS) fluctuations of the local energy.
The $3z$ GS basis yields identical results, within statistical error, to the $5z$ BFD basis.

Carbon atom DMC results for each basis set are shown in Table \ref{tab:c_atom_dmc}.
These calculations were performed with a time step of $\tau=0.01$ H$^{-1}$ which leads to negligible time step error for these high quality wavefunctions.
DMC depends less on basis size than VMC, as is immediately apparent from the data.
Nevertheless, both the G and GS basis sets outperform the BFD basis set.
The $3z$ GS basis yields identical results, within statistical error, to the $5z$ BFD basis, and an energy $0.1$ mH lower than the $3z$ BFD basis.

Both VMC and DMC results indicate that reoptimizing primitive basis function exponents provides improvements which can be significant for the GS basis.
In large systems, the ability to use a $3z$ basis in place of a $4z$ or $5z$ basis determines whether a calculation can be performed.
\begin{table}[htp]
  \begin{center}
    \caption{VMC energy and RMS fluctuations of the local energy, $\sigma$, in Hartrees for CAS(4,13) ground state of carbon using BFD, G, and GS basis functions.
      Statistical errors on the last digit are shown in parentheses.
      For each $n$, the $nz$ basis consists of $n$ $S$ functions, $n$ $P$ functions, and $n-1$ $D$ functions.
      CCSD(T) values for the BFD basis are included for comparison \cite{Burkatzki2007}.} \label{tab:c_atom}
    \begin{tabular}[b]{l|l|l|l}
      \hline\hline
      Type & Size    &Energy (H) & $\sigma$ (H)\\
      \hline
      BFD & $2z$   & -5.43161(3)  & 0.1395(6) \\
          & $3z$   & -5.43306(2)  & 0.099(3)  \\
          & $4z$   & -5.43332(2)  &  0.0904(2)\\
          & $5z$   & -5.43341(2)  & 0.0905(4) \\\hline
      G   & $2z$   & -5.43196(3)  & 0.138(2)  \\
          & $3z$   & -5.43324(2)  & 0.0989(5) \\\hline
      GS  & $2z$   & -5.43264(2)  & 0.1114(4) \\
          & $3z$   & -5.43344(2)  & 0.0898(2) \\ \hline
  CCSD(T) & $2z$   & -5.409230    & N/A \\
          & $3z$   & -5.427351    & N/A \\
          & $4z$   & -5.431486    & N/A \\
          & $5z$   & -5.432494    & N/A \\
      \hline\hline
    \end{tabular}
  \end{center}
\end{table}
\begin{table}[htp]
  \begin{center}
    \caption{DMC energy in Hartrees for CAS(4,13) ground state of carbon using BFD, G, and GS basis functions.
      Statistical errors on the last digit are shown in parentheses.
      For each $n$, the $nz$ basis consists of $n$ $S$ functions, $n$ $P$ functions, and $n-1$ $D$ functions.
      Calculations were performed with a time step of $\tau=0.01$ H$^{-1}$ which leads to a negligible time step error for these high quality wavefunctions.} \label{tab:c_atom_dmc}
    \begin{tabular}[b]{l|l|l}
      \hline\hline
      Type & Size    &Energy (H) \\
      \hline
      BFD & $2z$   & -5.43314(2)  \\
          & $3z$   & -5.43395(2)  \\
          & $4z$   & -5.43404(1)  \\
          & $5z$   & -5.43407(1)  \\\hline
      G   & $2z$   & -5.43342(2) \\
          & $3z$   & -5.43400(2) \\\hline
      GS  & $2z$   & -5.43356(2)  \\
          & $3z$   & -5.43407(1) \\
      \hline\hline
    \end{tabular}
  \end{center}
\end{table}
\subsection{Excited States of Carbon}
We consider the lowest lying excited states of carbon with $^5S^o$, $^3P^o$, $^1D^o$, and $^3F^o$ symmetries.
These states have configurations $2s^1 2p^3$, $2s^2 2p^1 3s^1$, $2s^2 2p^1 3d^1$, and $2s^2 2p^1 3d^1$, respectively.
The $^3P^o$, $^1D^o$, and $^3F^o$ states have much higher energy than the ground state and $^5S^o$ excited state.

The dominant CSF for each of these three states occupies orbitals that are unoccupied in the HF ground state.
For fair comparison, the BFD basis therefore must be augmented.
The diffuse functions of the aug-cc-pVnZ basis sets \cite{Kendall1992, Schuchardt2007, Feller1996} are used for this purpose.
The BFD $nz$ basis then becomes an $(n+1)z$ basis.

Obtained by application of the projection operators discussed in Section \ref{sec:c_atom}, the dominant CSF for each of the four excited states has one, one, four, and three Slater determinants, respectively.

\begin{figure}[htp]
 \begin{center}
   \includegraphics[scale=0.65]{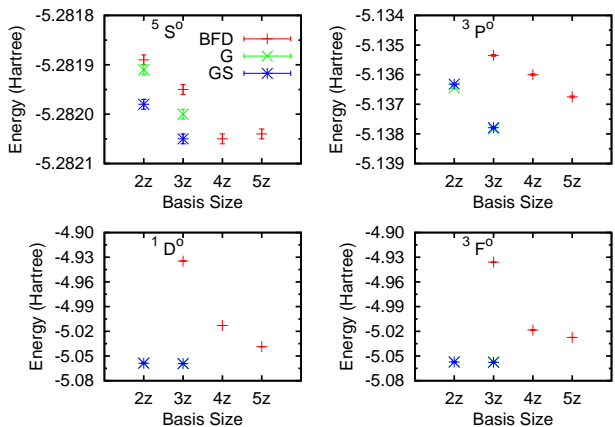}
   \caption{VMC energies in Hartrees for lowest lying excited states of carbon with $^5S^o$, $^3P^o$, $^1D^o$, $^3F^o$ symmetries.
   For each $n$, the $nz$ basis consists of $n$ $S$ functions, $n$ $P$ functions, and $n-1$ $D$ functions (where applicable).
   For $^3P^o$, $^1D^o$, $^3F^o$ calculations, the BFD basis is augmented with diffuse functions of the aug-cc-pVnZ basis sets \cite{Kendall1992, Schuchardt2007, Feller1996}.
   For G and GS basis sets, only $2z$ and $3z$ calculations were performed.
   In many cases, results for G and GS bases are indistinguishable on this scale.}
   \label{fig:energy_excited}
 \end{center}
\end{figure}

\begin{figure}[htp]
 \begin{center}
   \includegraphics[scale=0.65]{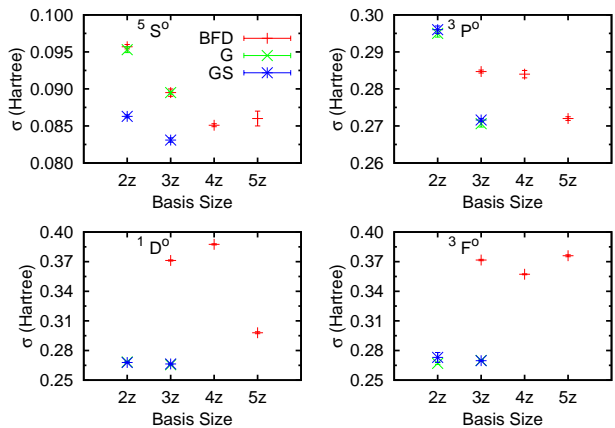}
   \caption{RMS fluctuations of VMC local energies. See Figure \ref{fig:energy_excited} for notation and details.}
   \label{fig:sigma_excited}
 \end{center}
\end{figure}

VMC results for energies and $\sigma$ of each system are shown in Figures \ref{fig:energy_excited} and \ref{fig:sigma_excited}.
In all cases, the reoptimized exponents provide significant gains in both energy and $\sigma$.
Results for the three higher lying states demonstrate that reoptimized exponents are essential for describing states containing orbitals unoccupied in the HF ground state.
For these systems, $2z$ results using the G and GS basis sets are substantially better than $5z$ results for the augmented BFD basis set.
In the most extreme case of $^3F^o$, the $2z$ GS basis results in 30 mH lower energy and 110 mH lower $\sigma$ than the $5z$ BFD basis.

The importance of the reoptimized exponents is evident for the excited states of carbon.
However, benefits of the GS basis relative to the G basis are never more than several tenths of a mH.
On the scale of the plots in Figures \ref{fig:energy_excited} and \ref{fig:sigma_excited}, many G and GS basis results coincide.
\subsection{Carbon Dimer}
Single determinant calculations were performed for the carbon dimer with initial wavefunctions generated from the QC code GAMESS \cite{Schmidt1993}.

VMC results for each basis set are shown in Table \ref{tab:c2}.
The G and GS basis sets outperform the BFD basis set.
In particular, the $2z$ G basis attains a 0.6 mH lower energy than the corresponding BFD basis, and the GS basis yields a 3.2 mH lower energy than the BFD basis.
The $3z$ GS basis yields an energy within 0.3 mH of and a $\sigma$ identical to the $5z$ BFD basis results.

DMC results for each basis set are shown in Table \ref{tab:c2_dmc}.
These calculations were performed with a time step of $\tau=0.005$ H$^{-1}$ which leads to negligible time step error.
The $2z$ G and GS basis sets significantly outperform the corresponding BFD basis set.
The $2z$ GS basis yields a result that is essentially converged with respect to basis size.

Both VMC and DMC results indicate that reoptimizing primitive basis function exponents provides improvements which can be significant for the GS basis.
\begin{table}[htp]
  \begin{center}
    \caption{VMC energy and $\sigma$ in Hartrees for single determinant carbon dimer ground state.
      Statistical errors on the last digit are shown in parentheses.
      For each $n$, the $nz$ basis consists of $n$ $S$ functions, $n$ $P$ functions, $n-1$ $D$ functions, $n-2$ $F$ functions, and $n-3$ $G$ functions.} \label{tab:c2}
    \begin{tabular}[b]{l|l|l|l}
      \hline\hline
      Type & Size    &Energy (H) & $\sigma$ (H)\\
      \hline
      BFD & $2z$   &  -11.02644(4)   & 0.4343(9) \\
          & $3z$   &  -11.03003(4)   & 0.4172(4) \\
          & $4z$   &  -11.03094(4)   & 0.4127(7) \\
          & $5z$   &  -11.03095(4)   & 0.4113(6) \\\hline
      G   & $2z$   &  -11.02707(4)   & 0.4288(7) \\
          & $3z$   &  -11.03030(4)   & 0.4183(6) \\\hline
      GS  & $2z$   &  -11.02968(4)   & 0.4191(6) \\
          & $3z$   &  -11.03065(4)   & 0.4109(6) \\
      \hline\hline
    \end{tabular}
  \end{center}
\end{table}
\begin{table}[htp]
  \begin{center}
    \caption{DMC energy in Hartrees for ground state of carbon dimer using BFD, G, and GS basis functions.
      Statistical errors on the last digit are shown in parentheses.
      For each $n$, the $nz$ basis consists of $n$ $S$ functions, $n$ $P$ functions, $n-1$ $D$ functions, $n-2$ $F$ functions, and $n-3$ $G$ functions.
      Calculations were performed with a time step of $\tau=0.005$ H$^{-1}$ which leads to negligible time step error.} \label{tab:c2_dmc}
    \begin{tabular}[b]{l|l|l}
      \hline\hline
      Type & Size    &Energy (H) \\
      \hline
      BFD & $2z$   &   -11.05561(3) \\
          & $3z$   &   -11.05719(4) \\
          & $4z$   &   -11.05728(4)\\
          & $5z$   &   -11.05723(4)\\ \hline
      G   & $2z$   &   -11.05632(3)\\
          & $3z$   &   -11.05717(3)\\\hline
      GS  & $2z$   &   -11.05702(4)\\
          & $3z$   &   -11.05719(3)\\
      \hline\hline
    \end{tabular}
  \end{center}
\end{table}
\subsection{Naphthalene}
Single determinant calculations were performed for naphthalene, C$_{10}$H$_8$, with initial wavefunctions generated from the QC code GAMESS \cite{Schmidt1993}.

Calculations were performed only for the $2z$ basis.
The intention of this section is not to produce an energy converged with respect to basis size, but to demonstrate that the utility of reoptimizing primitive basis functions, and GS primitives in particular, extends to large systems.
VMC and DMC results for each basis set are shown in Table \ref{tab:naphthalene}.
The DMC calculations were performed with a time step of $\tau=0.01$ H$^{-1}$.

At both the VMC and DMC level, wavefunction quality increases by reoptimizing the primitive Gaussian exponents.
The GS basis provides further improvement.
In particular, even at the DMC level, the $2z$ G basis attains a 4 mH lower energy than the corresponding BFD basis, and the GS basis yields a 15 mH lower energy than the BFD basis.
This is significant since DMC is less sensitive to basis set choice than VMC.
\begin{table}[htp]
  \begin{center}
    \caption{VMC energy and $\sigma$, and DMC energy in Hartrees for single determinant ground state naphthalene, C$_{10}$H$_8$.
      Statistical errors on the last digit are shown in parentheses.
      DMC calculations were performed with a time step of $\tau=0.01$ H$^{-1}$.
      For carbon, the $2z$ basis includes 2 $S$ function, 2 $P$ function, 1 $D$ function. 
      For hydrogen, the $2z$ basis includes 2 $S$ functions and 1 $P$ function.} \label{tab:naphthalene}
    \begin{tabular}[b]{l|l|l|l|l}
      \hline\hline
      Type & Size    &VMC Energy (H) &VMC $\sigma$ (H) &DMC Energy (H)\\
      \hline
      BFD & $2z$   &  -61.5193(5)  & 0.980(1) & -61.6479(5) \\ \hline
      G   & $2z$   &  -61.5273(4)  & 0.938(1) & -61.6518(5) \\\hline
      GS  & $2z$   &  -61.5438(4)  & 0.927(2) & -61.6634(5)  \\
      \hline\hline
    \end{tabular}
  \end{center}
\end{table}
\section{Conclusion}
\label{sec:conc}
Quantum Monte Carlo (QMC) methods have the advantage over standard quantum chemistry (QC) methods of rapid convergence with increasing basis size.
Basis-size dependence for pseudopotential calculations in QMC is further reduced by two basis set improvements introduced in this work.
Calculations for the ground state of carbon, the lowest lying $^5S^o$, $^3P^o$, $^1D^o$, $^3F^o$ excited states of carbon, carbon dimer, and naphthalene demonstrate the utility of our contribution.

First, we reoptimized the primitive basis function exponents for each system because the exponents of standard QC and QMC basis sets, such as the Burkatzki, Filippi and Dolg (BFD) basis, represent a compromise.
These standard exponents are designed to yield good energies for some range of chemical environments and excitation levels, but they cannot be optimal for all systems.
We have shown that reoptimizing primitive basis function exponents for each system yields significant improvements in the energy and fluctuations of the local energy, $\sigma$.
The most pronounced benefits were observed in higher-lying excited state calculations.
In the most extreme case of $^3F^o$ at the variational Monte Carlo (VMC) level, the $2z$ mixed basis was 30 mH lower in energy and 110 mH lower in $\sigma$ than the $5z$ numerical basis.
Although not discussed in this paper, we have found that reoptimization of standard Slater basis exponents used in all-electron calculations also provides considerable improvements in energy and $\sigma$.

Second, we introduced Gauss-Slater (GS) basis functions for non-divergent pseudopotential calculations.
GS functions behave like Gaussians at short distances and Slaters at long distances.
In all systems considered, results obtained using a mixed basis comprised of contracted and primitive basis functions improved when optimized Gaussian primitives were replaced by optimized GS primitives.
Importantly, for carbon dimer at the DMC level the $2z$ GS total energies are nearly converged with respect to basis size.

A $3z$ mixed basis with optimized GSs for carbon atom or carbon dimer produces results comparable to the $5z$ BFD basis.
Since the number of orbital coefficients to be optimized scales quadratically with basis size, the use of a more compact basis allows larger problems to be attacked in QMC.
\section{Acknowledgments}
This work was supported by the NSF (grant DMR-0908653) and by the DOE (grant DOE-DE-FG05-08OR23336).
Computations were performed in part at the Cornell NanoScale Facility, a member of the National Nanotechnology Infrastructure Network, and, at the Computation Center for Nanotechnology Innovation at Rensselaer Polytechnic Institute.
\bibliographystyle{apsrev}
\bibliography{gs}
\appendix
\section{Scaling Relations}
\label{sec:scaling}
To derive the scaling relation between $N_n^\zeta$ and $N_n^1$, consider
\begin{align}
1 &= \left(N_n^\zeta\right)^2 \int_0^\infty dr \; r^2 \; r^{2(n-1)} e^{-2\frac{(\zeta r)^2}{1+\zeta r}}\\
&= \frac{1}{\zeta^{2n+1}} \left(N_n^\zeta\right)^2 \int_0^\infty du \; u^{2n} e^{-2\frac{u^2}{1+u}} \\
&= \frac{1}{\zeta^{2n+1}} \left(\frac{N_n^\zeta}{N_n^1} \right)^2.
\end{align}
Hence, the scaling relation for the normalization factor is
\beq
N_n^\zeta = \zeta^{n+1/2} \; N_n^1. \label{eqn:norm_scaling}
\eeq
The values of $N_n^1$ are given in Table \ref{tab:norm}.
\begin{table}[htp]
  \begin{center}
    \caption{Normalization factors for Gauss-Slater basis functions with unit exponent and principal quantum number $n$.} \label{tab:norm}
    \begin{tabular}[b]{l|l}
      \hline\hline
      $n$ & $N_n^1$ \\
      \hline
      1 &  1.126467421\\
      2 &  0.576609950\\
      3 &  0.196581141\\
      4 &  0.050275655\\
      5 &  0.010280772\\
      \hline\hline
    \end{tabular}
  \end{center}
\end{table}

To derive the scaling relations for the parameters $\alpha_i^\zeta$ and $c_i^\zeta$ in the Gaussian expansion of the Gauss-Slater functions,
suppose the best-fit expansion for $\zeta=1$ is
\beq
N_n^1 \; e^{-\frac{r^2}{1+ r}} = \sum_i c_i^1 \;
\sqrt{\frac{2(2\alpha_i^1)^{n+\frac{1}{2}}}{\Gamma(n+\frac{1}{2})}} \; e^{-\alpha_i^1 r^2}.
\eeq
Using Eqn. (\ref{eqn:norm_scaling}) and performing the substitution $r \to \zeta r$ results in
\begin{align}
N_n^\zeta \; e^{-\frac{(\zeta r)^2}{1+\zeta r}} &= \zeta^{n+1/2} \sum_i c_i^1 \;
\sqrt{\frac{2(2\alpha_i^1)^{n+\frac{1}{2}}}{\Gamma(n+\frac{1}{2})}} \; e^{-\alpha_i^1 \zeta^2 r^2} \\
&=  \sum_i c_i^1 \;
\sqrt{\frac{2(2\alpha_i^1 \zeta^2)^{n+\frac{1}{2}}}{\Gamma(n+\frac{1}{2})}} \; e^{-\alpha_i^1 \zeta^2 r^2} \\
&= \sum_i c_i^\zeta \;
\sqrt{\frac{2(2\alpha_i^\zeta)^{n+\frac{1}{2}}}{\Gamma(n+\frac{1}{2})}} \; e^{-\alpha_i^\zeta r^2},
\end{align}
where
\begin{align}
\alpha_i^\zeta &= \zeta^2 \alpha_i^1 \\
c_i^\zeta &= c_i^1.
\end{align}
\section{Spatial Derivatives}
A general unnormalized radial basis function has the form
\begin{equation}
R^\zeta_n(r)  = r^{n-1} e^{g^\zeta(r)},
\end{equation}
where $g^\zeta(r)$ is an arbitrary function.
The gradient is
\beq
\bm{\nabla} R^{\zeta}_n(r) =  \D{R^{\zeta}_n(r)}{r} \hat{\bm{r}},
\eeq
where
\beq
\D{R^{\zeta}_n(r)}{r} =  R^{\zeta}_n(r)\left[\frac{(n-1)}{r}+\D{g^\zeta(r)}{r}\right].
\eeq
The Laplacian is
\beq
\nabla^2 R^{\zeta}_n(r) = \D{^2 R^{\zeta}_n(r)}{r^2} + \frac{2}{r}\D{R^{\zeta}_n(r)}{r},
\eeq
where
\begin{align}
\D{^2R^{\zeta}_n(r)}{r^2} &=   R^{\zeta}_n(r) \left[\D{^2 g^\zeta(r)}{r^2} - \frac{(n-1)}{r^2}\right] \notag \\
&\qquad+ \frac{1}{R^{\zeta}_n(r)} \left(\D{R^{\zeta}_n(r)}{r}\right)^2 .
\end{align}
For Gauss-Slater functions,
\begin{align}
g^\zeta(r) &= -\frac{(\zeta r)^2}{1+\zeta r} \\
\D{g^\zeta(r)}{r} &= - \frac{r\zeta^2(2+\zeta r)}{(1+\zeta r)^2}\\
\D{^2g^\zeta(r)}{r^2} &= - \frac{2 \zeta^2}{(1+\zeta r)^3}.
\end{align}
For Gaussian functions,
\begin{align}
g^\zeta(r) &= -\zeta r^2 \\
\D{g^\zeta(r)}{r} &= -2 \zeta r \\
\D{^2g^\zeta(r)}{r^2} &= - 2 \zeta.
\end{align}
For Slater functions,
\begin{align}
g^\zeta(r) &= -\zeta r \\
\D{g^\zeta(r)}{r} &= -\zeta \\
\D{^2g^\zeta(r)}{r^2} &= 0.
\end{align}
\section{Parameter Derivatives}
Wavefunction optimization via the linear method requires both the derivatives of the wavefunction with respect to the exponent parameters $\zeta$, and the Hamiltonian acting on those derivatives.
From Eqn. (\ref{eqn:norm_scaling}), the derivative of the normalization with respect to the exponent is
\beq
\D{N_n^\zeta}{\zeta} = \frac{(n+1/2)}{\zeta} N_n^\zeta.
\eeq
Now consider a general unnormalized radial basis function of the form
\beq
R^{\zeta}_n(r) = r^{n-1} e^{g^\zeta(r)},
\eeq
where $g^\zeta(r)$ is an arbitrary function.
The derivative of the radial part of the wavefunction with respect to the exponent is
\begin{align}
\D{R^{\zeta}_n(r)}{\zeta}  = f^\zeta(r) R^{\zeta}_n(r),
\end{align}
where
\beq
f^\zeta(r) \equiv \D{g^\zeta(r)}{\zeta}.
\eeq
The gradient is
\beq
\bm{\nabla}\left[\D{R^{\zeta}_n(r)}{\zeta}\right] = \D{f^\zeta(r)}{r} R^{\zeta}_n(r) \hat{\bm{r}} + f^\zeta(r) \left[\bm{\nabla} R^{\zeta}_n(r)\right].
\eeq
The Laplacian is
\begin{align}
\nabla^2\left[\D{R^{\zeta}_n(r)}{\zeta}\right] &= \frac{2}{r} \D{f^\zeta(r)}{r} \left[R^{\zeta}_n(r) +  r\D{R^{\zeta}_n(r)}{r} \right]  \notag \\
&+\D{^2 f^\zeta(r)}{r^2} R^{\zeta}_n(r) + f^\zeta(r) \nabla^2  R^{\zeta}_n(r).
\end{align}
For Gauss-Slater functions,
\begin{align}
f^\zeta(r) &= -\frac{  \zeta r^2 ( 2+\zeta r) } {(1+\zeta r)^2} \\
\D{f^\zeta(r)}{r} &= -\frac{  \zeta r [ 4+\zeta r (3+\zeta r)] } {(1+\zeta r)^3} \\
\D{^2 f^\zeta(r)}{r^2} &= \frac{ 2 \zeta  (\zeta r - 2) } {(1+\zeta r)^4}.
\end{align}
For Gaussian functions,
\begin{align}
f^\zeta(r) &= -r^2 \\
\D{f^\zeta(r)}{r} &= -2r \\
\D{^2 f^\zeta(r)}{r^2} &=  -2.
\end{align}
For Slater functions,
\begin{align}
f^\zeta(r) &= -r \\
\D{f^\zeta(r)}{r} &= -1 \\
\D{^2 f^\zeta(r)}{r^2} &=  0.
\end{align}
\section{Exponents}
To promote use of this basis, exponents for each system studied are given in this appendix.
Only exponents of the primitives are given, as the contracted functions are presented elsewhere \cite{Burkatzki2007}.
Exponents for the ground state of carbon using CAS(4,13) CSFs, and, the ground state of carbon dimer using a single CSF are shown in Table \ref{tab:c_exp}.
\begin{table}[htp]
  \begin{center}
    \caption{
      Basis exponents for CAS(4,13) ground state of carbon and ground state of carbon dimer using G and GS basis functions.
      For each $n$, the $nz$ basis includes $n-1$ $S$ primitives, $n-1$ $P$ primitives, $n-1$ $D$ primitives, and $n-2$ $F$ primitives (where applicable).} \label{tab:c_exp}
    \begin{tabular}[b]{l|l|l|l|l}
      \hline\hline
      Type & Size    & $L$ & C Exp. & C$_2$ Exp.\\
      \hline
      G   & $2z$   &  $S$   & 0.087 &  0.145  \\
          &        &  $P$   & 0.129 &  0.196\\
          &        &  $D$   & 0.470 &  0.679\\ \cline{2-5}
          & $3z$   &  $S$   & 0.102 &  0.127\\
          &        &  $S$   & 0.676 &  0.998\\
          &        &  $P$   & 0.104 &  0.121\\ 
          &        &  $P$   & 0.270 &  0.423\\ 
          &        &  $D$   & 0.314 &  0.386\\ 
          &        &  $D$   & 0.982 &  1.099\\ 
          &        &  $F$   & N/A   &  0.783\\ \hline
      GS  & $2z$   &  $S$   & 0.586 &  0.853\\
          &        &  $P$   & 0.984 &  1.162\\
          &        &  $D$   & 1.810 &  3.774\\ \cline{2-5}
          & $3z$   &  $S$   & 1.000 &  1.127\\
          &        &  $S$   & 1.258 &  1.570\\
          &        &  $P$   & 1.059 &  0.703\\ 
          &        &  $P$   & 1.750 &  1.416\\ 
          &        &  $D$   & 1.132 &  2.225\\ 
          &        &  $D$   & 1.981 &  3.228\\
          &        &  $F$   & N/A   &  2.588\\ \hline
      \hline
    \end{tabular}
  \end{center}
\end{table}
Exponents for the lowest lying excited states of carbon with $^5S^o$, $^3P^o$, $^1D^o$, $^3F^o$ symmetries are shown in Table \ref{tab:c_excited_exp}.
\begin{table}[htp]
  \begin{center}
    \caption{
      Basis exponents for the lowest lying excited states of carbon with $^5S^o$, $^3P^o$, $^1D^o$, $^3F^o$ symmetries.
      For each $n$, the $nz$ basis includes $n-1$ $S$ primitives, $n-1$ $P$ primitives, and $n-1$ $D$ primitives (where applicable).} \label{tab:c_excited_exp}
    \begin{tabular}[b]{l|l|l|l|l|l|l}
      \hline\hline
      Type & Size  & $L$    & $^5S^o$ Exp.   & $^3P^o$ Exp.  & $^1D^o$ Exp. & $^3F^o$ Exp. \\
      \hline
      G   & $2z$   &  $S$   & 0.110     & 0.006    & 0.087   & 0.112    \\
          &        &  $P$   & 0.150     & 0.157    & 0.109   & 0.766    \\
          &        &  $D$   & N/A       & N/A      & 0.006   & 0.006    \\ \cline{2-7}
          & $3z$   &  $S$   &  0.356    & 0.010    & 0.094   & 0.096    \\
          &        &  $S$   &  2.145    & 0.284    & 0.496   & 0.689    \\
          &        &  $P$   &  0.809    & 4.868    & 0.079   & 0.117    \\ 
          &        &  $P$   &  2.262    & 6.707    & 0.854   & 0.524    \\ 
          &        &  $D$   &  N/A      & N/A      & 0.007   & 0.007    \\ 
          &        &  $D$   &  N/A      & N/A      & 0.407   & 0.822    \\ \hline
      GS  & $2z$   &  $S$   & 1.960     & 0.059    & 0.756   & 0.570     \\
          &        &  $P$   & 1.860     & 3.002    & 0.763   & 0.588   \\
          &        &  $D$   & N/A       & N/A      & 0.351   & 0.293   \\ \cline{2-7}
          & $3z$   &  $S$   & 1.148     &  0.371   & 0.160   & 0.894    \\
          &        &  $S$   & 1.451     &  0.376   & 0.931   & 2.045    \\
          &        &  $P$   & 1.137     &  0.691   & 0.043   & 1.326    \\ 
          &        &  $P$   & 1.325     &  2.087   & 0.444   & 3.016    \\ 
          &        &  $D$   & N/A       &  N/A     & 0.238   & 0.295      \\ 
          &        &  $D$   & N/A       &  N/A     & 0.953   & 1.545   \\ \hline
      \hline
    \end{tabular}
  \end{center}
\end{table}
Exponents for the ground state of naphthalene are shown in Table \ref{tab:nap_exp}.
In naphthalene, since atoms of the same atomic species are located at inequivalent geometrical locations,
one could independently optimize the exponents for each inequivalent atom, but we have
not done so because we expect the resulting gain to be small.

We have found that the carbon $S$ and $P$ exponents change relatively little from one molecule
to another (though they do differ more for the atom) while there is considerable leeway in the
$D$ exponents (they change considerably even from one optimization to another for a given molecule).
This is because the energy and $\sigma$ are not as sensitive to the $D$ basis functions as they do not appear in the ground-state determinant of the carbon atom.
Hence it is possible to find an approximately optimal set of exponents for the atoms in a large molecule,
by optimizing them for a small molecule with the same atoms.
\begin{table}[htp]
  \begin{center}
    \caption{
      Basis exponents for ground state of naphthalene, C$_{10}$H$_8$, using G and GS basis functions.
      For carbon, the $2z$ basis includes 1 $S$ primitive, 1 $P$ primitive, 1 $D$ primitive. 
      For hydrogen, the $2z$ basis includes 1 $S$ primitive.} \label{tab:nap_exp}
    \begin{tabular}[b]{l|l|l|l|l}
      \hline\hline
      Type & Size    & $L$ & C$_{10}$H$_8$ C Exp. & C$_{10}$H$_8$ H Exp.\\
      \hline
      G   & $2z$   &  $S$   & 0.139 & 0.099  \\
          &        &  $P$   & 0.191 & N/A \\
          &        &  $D$   & 0.754 & N/A \\ \cline{2-5}
      GS  & $2z$   &  $S$   & 0.875 & 0.798 \\
          &        &  $P$   & 1.118 & N/A \\
          &        &  $D$   & 2.109 & N/A \\ \hline
      \hline
    \end{tabular}
  \end{center}
\end{table}

\end{document}